\documentclass{emulateapj}
\usepackage[]{natbib, graphicx}

\begin{document}

\title{HST WFC3/IR Observations of Active Galactic Nucleus Host Galaxies at $z\sim2$: Supermassive Black Holes Grow in Disk Galaxies$^1$}

\shorttitle{Black Holes Grow in Disks}
\shortauthors{Schawinski et al.}
\slugcomment{To appear in the Astrophysical Journal Letters}

\author{
Kevin Schawinski\altaffilmark{2,3,8}, Ezequiel Treister\altaffilmark{4,10}, C. Megan Urry\altaffilmark{2,3,5} , Carolin N. Cardamone\altaffilmark{6}, Brooke Simmons\altaffilmark{3,5} and Sukyoung K. Yi\altaffilmark{7}
}

\altaffiltext{1}{Based on observations made with the NASA/ESA Hubble Space Telescope, obtained from the data archive at the Space Telescope Institute. STScI is operated by the association of Universities for Research in Astronomy, Inc. under the NASA contract NAS 5-26555.}
\altaffiltext{2}{Department of Physics, Yale University, New Haven, CT 06511, U.S.A.}
\altaffiltext{3}{Yale Center for Astronomy and Astrophysics, Yale University, P.O. Box 208121, New Haven, CT 06520, U.S.A.}
\altaffiltext{4}{Institute for Astronomy, 2680 Woodlawn Drive, University of Hawaii, Honolulu, HI 96822, U.S.A.}
\altaffiltext{5}{Department of Astronomy, Yale University, New Haven, CT 06511, U.S.A.}
\altaffiltext{6}{Department of Physics, Massachusetts Institute of Technology, 77 Massachusetts Avenue, Cambridge, MA 02139}
\altaffiltext{7}{Department of Astronomy, Yonsei University, Seoul 120-749, Korea}
\altaffiltext{8}{Einstein Fellow}
\altaffiltext{9}{Chandra Fellow}

\email{kevin.schawinski@yale.edu}

\def\Chandra{\textit{Chandra}}
\def\XMM{\textit{XMM-Newton}}
\def\Swift{\textit{Swift}}

\def\OI{[\mbox{O\,{\sc i}}]~$\lambda 6300$}
\def\OIII{[\mbox{O\,{\sc iii}}]~$\lambda 5007$}
\def\SII{[\mbox{S\,{\sc ii}}]~$\lambda \lambda 6717,6731$}
\def\NII{[\mbox{N\,{\sc ii}}]~$\lambda 6584$}

\def\Ha{{H$\alpha$}}
\def\Hb{{H$\beta$}}

\def\NIIHa{[\mbox{N\,{\sc ii}}]/H$\alpha$}
\def\SIIHa{[\mbox{S\,{\sc ii}}]/H$\alpha$}
\def\OIHa{[\mbox{O\,{\sc i}}]/H$\alpha$}
\def\OIIIHb{[\mbox{O\,{\sc iii}}]/H$\beta$}

\def\Ebmv{E($B-V$)}
\def\LOIII{$L[\mbox{O\,{\sc iii}}]$}
\def\Ledd{${L/L_{\rm Edd}}$}
\def\LOIIIs4{$L[\mbox{O\,{\sc iii}}]$/$\sigma^4$}
\def\LOIIIMbh{$L[\mbox{O\,{\sc iii}}]$/$M_{\rm BH}$}
\def\Mbh{$M_{\rm BH}$}
\def\Msigma{$M_{\rm BH} - \sigma$}
\def\Ms{$M_{\rm *}$}
\def\Msun{$M_{\odot}$}
\def\Msunyr{$M_{\odot}yr^{-1}$}

\def\ergs{$~\rm erg~s^{-1}$}
\def\kms{$~\rm km~s^{-1}$}

\def\galfit{\texttt{GALFIT}}
\def\multidrizzle{\texttt{multidrizzle}}

\def\sersic{S\'{e}rsic}

\begin{abstract}
We present the rest-frame optical morphologies of active galactic nucleus (AGN) host galaxies at $1.5<z<3$, using near-infrared imaging from the \textit{Hubble Space Telescope} Wide Field Camera 3, the first such study of AGN host galaxies at these redshifts. The AGN are X-ray selected from the \textit{Chandra} Deep Field South and have typical luminosities of $10^{42} < L_{\rm X} < 10^{44}$\ergs. Accreting black holes in this luminosity and redshift range account for a substantial fraction of the total space density and black hole mass growth over cosmic time; they thus represent an important mode of black hole growth in the universe. We find that the majority ($\sim80\%$) of the host galaxies of these AGN have low \sersic\ indices indicative of disk-dominated light profiles, suggesting that secular processes govern a significant fraction of the cosmic growth of black holes. That is, many black holes in the present-day universe grew much of their mass in disk-dominated galaxies and not in early-type galaxies or major mergers. The properties of the AGN host galaxies are furthermore indistinguishable from their parent galaxy population and we find no strong evolution in either effective radii or morphological mix between $z\sim2$ and $z\sim0.05$.
\end{abstract}

\keywords{galaxies: Seyfert; galaxies: high-redshift; galaxies: active}

\section{Introduction}
\label{sec:intro}
Host galaxy morphology is a key parameter in the joint formation of galaxies and supermassive black holes via active black hole growth phases \citep{2010ApJ...711..284S}. In the local universe, black hole growth is associated with very different evolutionary stages of galaxies: early-type host galaxies feature black hole accretion in a specific time window after a rapidly suppressed burst of star formation that was induced by a merger event \citep{2007MNRAS.382.1415S, 2010ApJ...714L.108S, 2010ApJ...711..284S, 2007MNRAS.381..543W, 2008ApJ...673..715C}, while black hole growth in late-type galaxies, which dominates active galactic nucleus (AGN) host galaxy sample by number, are massive, stable disk galaxies with no obvious recent perturbations to star formation \citep{2010ApJ...711..284S}.  That is major mergers do not seem to make up a large part of the AGN host galaxy population at low redshift, although puzzlingly the fraction of AGN exhibiting signs of recent or ongoing interactions increases at the highest luminosities in unbiased, hard X-ray selected samples such as the \Swift\ BAT sample \citep{2009ApJ...692L..19S, 2010ApJ...716L.125K}. In any case, accretion at $z\sim0$ represent a negligible fraction of cosmic black hole growth, most of which occurred at high redshift.

\begin{figure*}
\begin{center}

\includegraphics[angle=90,width=0.163\textwidth]{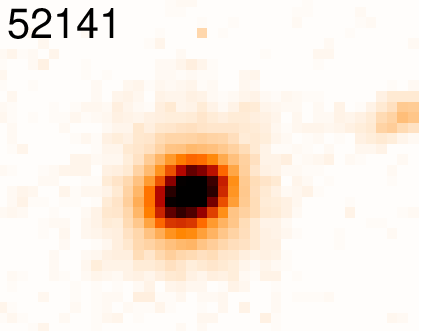}
\includegraphics[angle=90,width=0.163\textwidth]{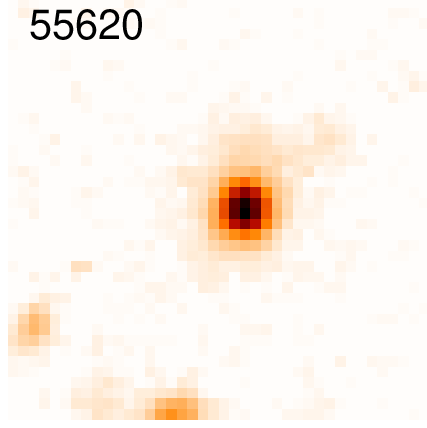}
\includegraphics[angle=90,width=0.163\textwidth]{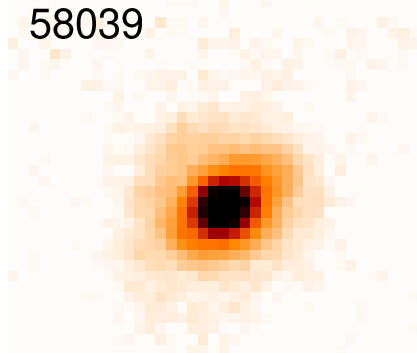}\\
\includegraphics[angle=90,width=0.163\textwidth]{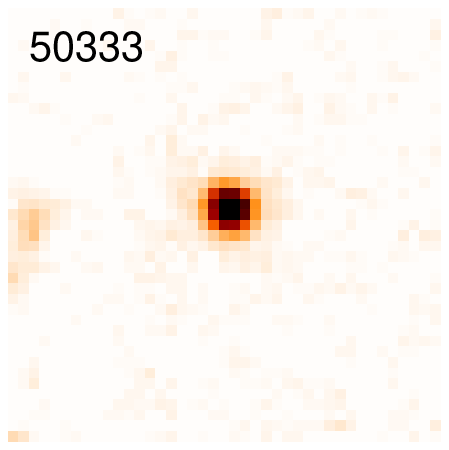}
\includegraphics[angle=90,width=0.163\textwidth]{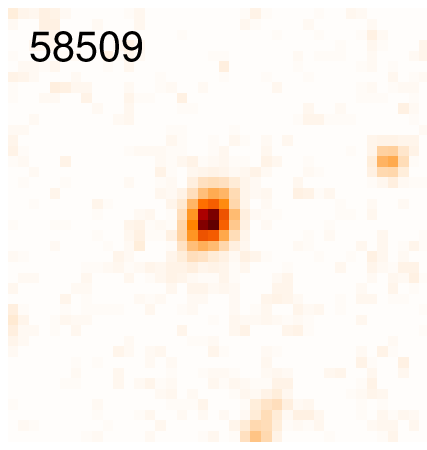}
\includegraphics[angle=90,width=0.163\textwidth]{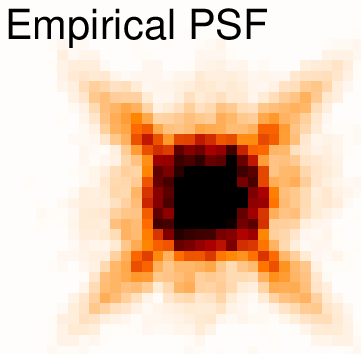}

\caption{Sample $H$-band (F160W) cutouts measuring $5.6\arcsec \times 5.6\arcsec$ of the X-ray selected AGN host galaxies in our sample. In the top-left of each cutout, we give the corresponding ID number (see Table \ref{tab:agn}). In the top row are galaxy-dominated AGN host galaxies while the left and middle of the bottom row are AGN-dominated sources. In the bottom-right panel, we show the empirical PSF generated from stars in the field.   \label{fig:gallery}}

\end{center}
\end{figure*}

The properties of AGN host galaxies during the peak epoch of both star formation and black hole growth at $z\sim2$ have until very recently remained virtually inaccessible. Ground-based imaging does not offer sufficient angular resolution to resolve the AGN host galaxies, while most \textit{Hubble Space Telescope} imaging was in optical bands that translates to the rest-frame ultraviolet at $z\sim2$. With the installation of the new Wide Field Camera 3 (WFC3) on the \textit{Hubble Space Telescope}, we now have available ultra-deep, high resolution near-infrared images of AGN at $z\sim2$ which allow us to study the \textit{rest-frame optical} properties of their host galaxies in detail. The $F160W$ ($H$-band) filter corresponds approximately to the $V$-band at $z\sim2$, with spatial resolution comparable to or better than Sloan Digital Sky imaging at $z\sim0.05$\footnote{For SDSS, the $1.^{''}4$ median seeing at $z=0.05$ corresponds to 1.36 kpc, while for \textit{HST} WFC3/IR, the $0.^{''}13$ (undersampled) pixel scale at $z=2$ corresponds to 1.09 kpc.}.

In this \textit{Letter}, we present the  restframe optical morphologies of moderate luminosity AGN ($10^{42} < L_{\rm X} < 10^{44}$\ergs; corresponding to $-23 \lesssim M_{\rm V} \lesssim -18$) during the peak epoch of growth at $z\sim2$. AGN with these luminosities represent a significant fraction of the cosmic black hole growth in terms of both number density and in X-ray light emitted \citep[e.g.,][]{2003ApJ...598..886U, 2005A&A...441..417H}.  Throughout this \textit{Letter}, we assume a $\Lambda$CDM cosmology with $h_{0}=0.7$, $\Omega_{m}=0.27$ and $\Omega_{\Lambda}=0.73$, in agreement with the most recent cosmological observations \citep{2009ApJS..180..225H}.

\begin{figure}
\begin{center}

\includegraphics[angle=90, width=0.49\textwidth]{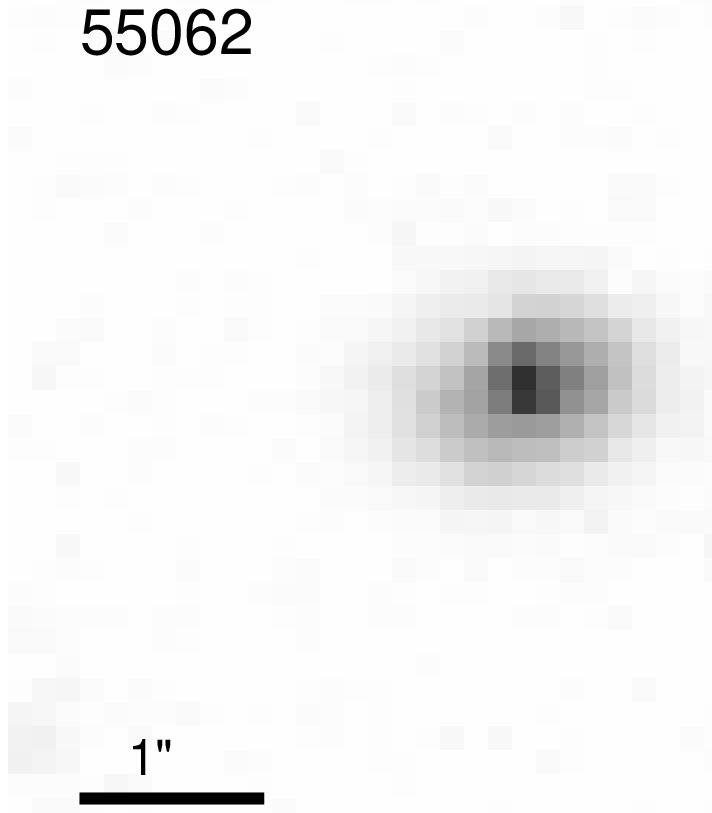}\\
\includegraphics[angle=90, width=0.49\textwidth]{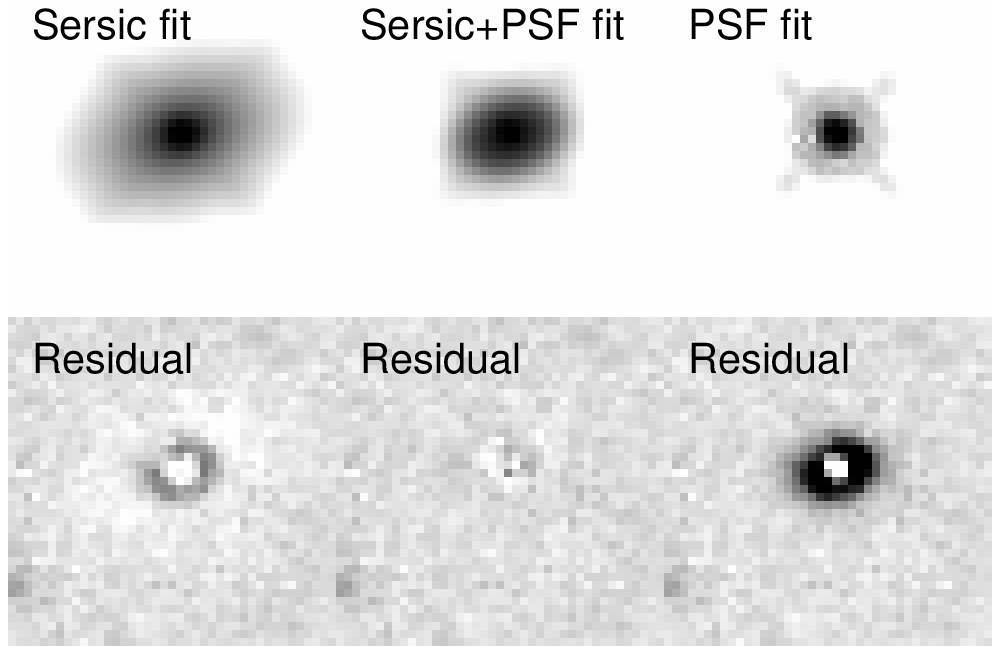}
\caption{An example of an F160W images of AGN host galaxies and the \galfit\ output for a galaxy dominated by starlight but containing a nuclear point source. Below the image are the three best-fit models (\sersic\ only, \sersic+PSF and PSF only), with a logarithmic stretch, and below that are the corresponding residual images. The PSF-only residuals clearly show a resolved component that the PSF-only fit could not accommodate. \label{fig:examples}}

\end{center}
\end{figure}

\begin{deluxetable*}{lcccccccc}
\tablecolumns{9}
\tablewidth{0pc}
\tabletypesize{\scriptsize}
\tablecaption{X-ray-selected AGN host galaxies in the WFC3/IR ERS field}
\tablehead{
 \colhead{ID$^1$} & 
 \colhead{RA} & 
 \colhead{Dec} & 
 \colhead{Redshift} & 
 \colhead{Type$^2$} &
 \colhead{$log_{10}(L_{\rm{X, obs}})$} &
 \colhead{$log_{10}(M_{\rm{stellar}})$$^3$} &
 \colhead{Eddington} &
 \colhead{$M_{\rm V}$} \\
 \colhead{} & 
 \colhead{(J2000)} & 
 \colhead{(J2000)} & 
 \colhead{} & 
 \colhead{} &
 \colhead{\ergs} &
 \colhead{\Msun} &
 \colhead{Ratio$^4$} &
 \colhead{AB mag} 
}
\startdata
49190  & 03 32 19.9 &-27 45 17.9  & 2.424  & phot  & 44.05  & 10.24  & 0.84 & -22.17 \\
50057  & 03 32 06.7 &-27 44 55.1  & 2.296  & phot  & 42.94  & 10.25  & 0.06  & -22.05\\
50333  & 03 32 03.0 &-27 44 50.0  & 2.573  & spec  & 44.06  & 10.39  & 0.57  &-22.79\\
50634  & 03 32 04.0 &-27 44 41.5  & 3.618  & phot  & 43.52  & 10.23  & 0.25  &-22.85\\
52141  & 03 32 10.9 &-27 44 14.9  & 1.613  & spec  & 44.35  & 10.08  & 2.60  &-22.97\\
52399  & 03 32 14.8 &-27 44 02.6  & 1.527  & phot  & 42.78  & 10.32  & 0.04  &-19.95\\
53849  & 03 32 15.1 &-27 43 35.3  & 1.691  & phot  & 43.02  & 10.81  & 0.02  & -23.16\\
54369  & 03 32 01.6 &-27 43 27.0  & 2.720  & spec  & 44.47  & 10.41  & 1.38  & -22.23\\
55062  & 03 32 25.7 &-27 43 05.7  & 2.291  & spec  & 44.39  & 10.89  & 0.30  & -22.85\\
55620  & 03 32 24.2 &-27 42 57.7  & 2.303  & spec  & 43.39  & 10.54  & 0.08  & -22.57\\
56112  & 03 32 20.0 &-27 42 43.6  & 2.733  & phot  & 43.86  & 10.07  & 0.87  &  -21.35\\
56769  & 03 32 20.2 &-27 42 27.2  & 2.773  & phot  & 43.58  & \nodata  & \nodata &  -21.06\\
56954  & 03 32 14.1 &-27 42 30.1  & 2.026  & spec  & 42.99  & 10.49  & 0.04  &  -22.07\\
57420  & 03 32 25.2 &-27 42 18.8  & 1.617  & spec  & 44.00  & 11.05  & 0.08  &  -23.72\\
57805  & 03 32 15.0 &-27 42 25.0  & 1.895  & phot  & 43.80  & 11.48  & 0.02  & -25.13\\
57859  & 03 32 15.8 &-27 42 07.6  & 2.779  & phot  & 43.32  & 10.71  & 0.04  & -22.94\\
58039  & 03 32 33.9 &-27 42 04.1  & 1.936  & phot  & 42.85  & 10.85  & 0.01  &-23.57\\
58224  & 03 32 15.2 &-27 41 58.6  & 2.402  & spec  & 43.47  & 10.56  & 0.09  & -22.67\\
58330  & 03 32 05.0 &-27 42 02.7  & 2.062  & phot  & 43.16  & 10.40  & 0.07  &-23.29\\
58509  & 03 32 12.7 &-27 41 49.0  & 1.490  & phot  & 43.02  & 9.91  & 0.19  & -20.23\\
58657  & 03 32 08.3 &-27 41 53.5  & 2.470  & spec  & 43.68  & 10.35  & 0.26  &-21.68\\
59060  & 03 32 17.1 &-27 41 37.0  & 2.193  & phot  & 43.85  & 10.57  & 0.21  & -21.79\\
63732  & 03 32 35.4 &-27 40 02.7  & 1.490  & phot  & 42.60  & 10.27  & 0.03  &-22.13\\
\enddata

\label{tab:agn}

\tablenotetext{1}{Catalog ID, see \cite{2010ApJS..189..270C}.}
\tablenotetext{2}{Type of redshift: phot - photometric redshift, spec - spectroscopic redshift; see \cite{2010ApJS..189..270C}.}
\tablenotetext{3}{Stellar mass as computed by \cite{2010ApJ...721L..38C}.}
\tablenotetext{4}{Black hole mass is calculated from this stellar mass following the relation of \cite{2004ApJ...604L..89H}. Eddington ratios assume an X-ray to bolometric correction factor of 20. Since a substantial fraction of the total stellar mass is likely in a disk, the black hole masses are overestimates, which means the Eddington ratios are lower limits. }

\end{deluxetable*}

\section{Data}

\subsection{Sample Selection}
\label{sec:sample}
We select AGN using X-ray data from the deep 2 Ms \Chandra\ Deep Field South observations \citep{2008ApJS..179...19L} in the WFC3 Early Release Science (ERS) field covered by the near-infrared mosaic \citep{2010arXiv1005.2776W}. The  X-ray emission is a signpost for the presence of an accreting black hole.  The K correction shifts the hard X-rays, where obscuration becomes unimportant, to the observed \Chandra\ band, making X-ray selection highly efficient except for the most highly obscured, Compton-thick systems \citep[e.g.,][]{2009ApJ...706..535T}. The flux limit of the 2 Ms \Chandra\ data at $z\sim2$ means that all our X-ray sources are more luminous than $L_{\rm X} \sim 10^{42}$\ergs\ and thus are unlikely to be affected significantly by X-ray emission from star formation.

We identify the matched counterparts of the X-ray sources using the MUSYC catalog of \cite{2010ApJS..189..270C}, which includes either spectroscopic redshifts where available or highly accurate photometric redshifts derived from the combined broad- and medium-band photometry (\citealt{2010ApJS..189..270C}; photo-z quality $q_{z} < 3$). A total of 23 X-ray AGN, with WFC3/IR coverage and $z > 1.45$, constitute the sample investigated in this \textit{Letter}. We list the basic properties of the X-ray selected AGN host galaxies in Table \ref{tab:agn} and show a selection of $H$-band cutouts in Figure \ref{fig:gallery}. The mean and median redshifts of this sample are 2.21 and 2.29, respectively, with a standard deviation of 0.52. The highest redshift is 3.62. The mean and median observed X-ray luminosities are $10^{43.1}$  and $10^{43.3}$\ergs, respectively, with a standard deviation of 1.41 dex.

\subsection{WFC3/IR Image Reduction}
We retrieved the WFC3/IR $H$-band (F160W) images taken as part of the ERS program from the archive and processed them with the standard STSDAS \texttt{pyraf} task \multidrizzle\ to create a mosaicked F160W image of the entire ERS field. Using \multidrizzle, we also created a noise image from the inverse variance map for each pixel, which takes into account the noise sources in each exposure, including read-noise, dark current and sky background, as well as Poisson noise from the sources themselves and dependence on exposure time. We then created cutouts for each X-ray source from both the $H$-band science and noise images for further analysis.

\begin{figure*}
\begin{center}

\includegraphics[angle=90, width=0.48\textwidth]{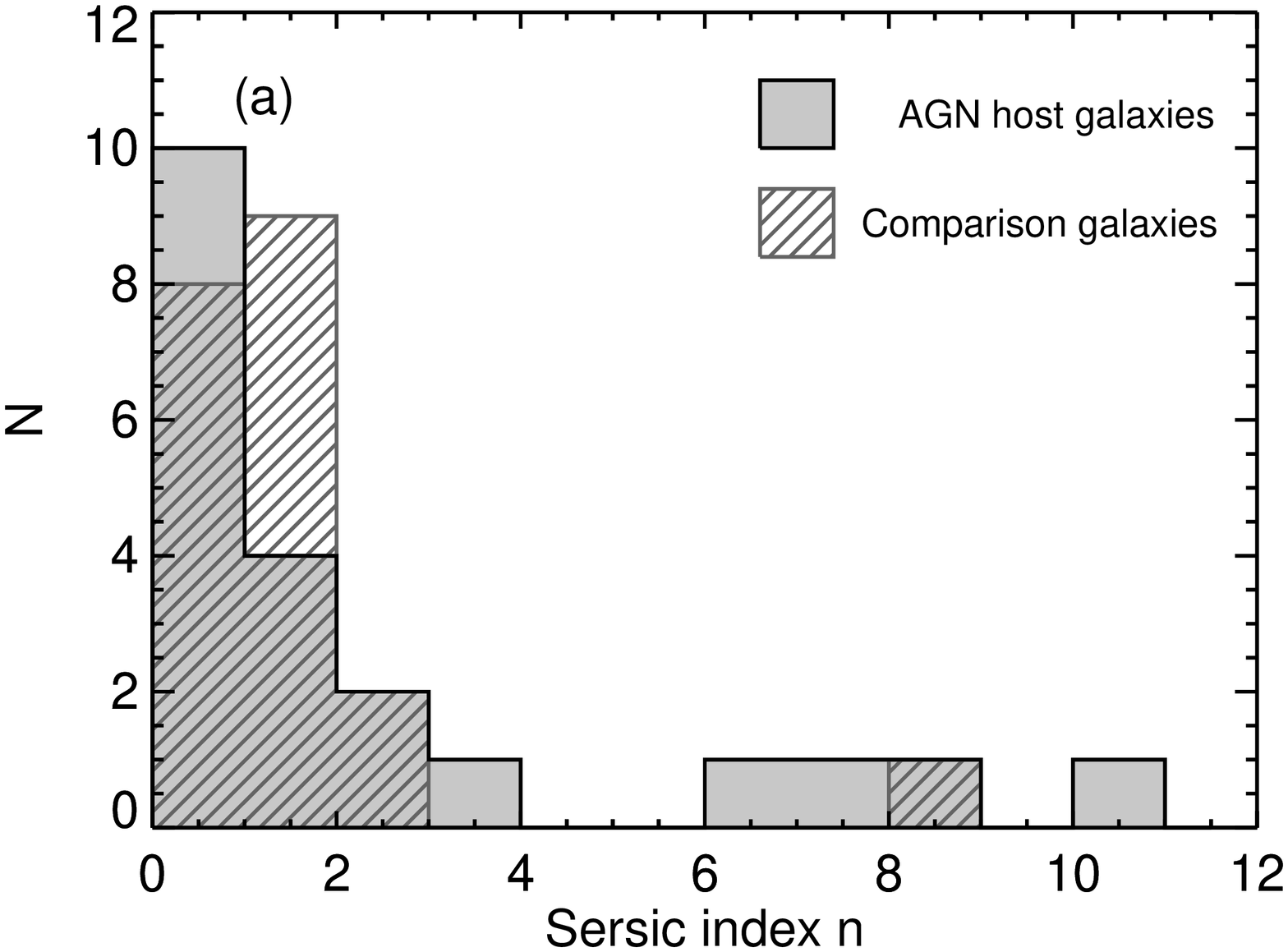}
\includegraphics[angle=90, width=0.48\textwidth]{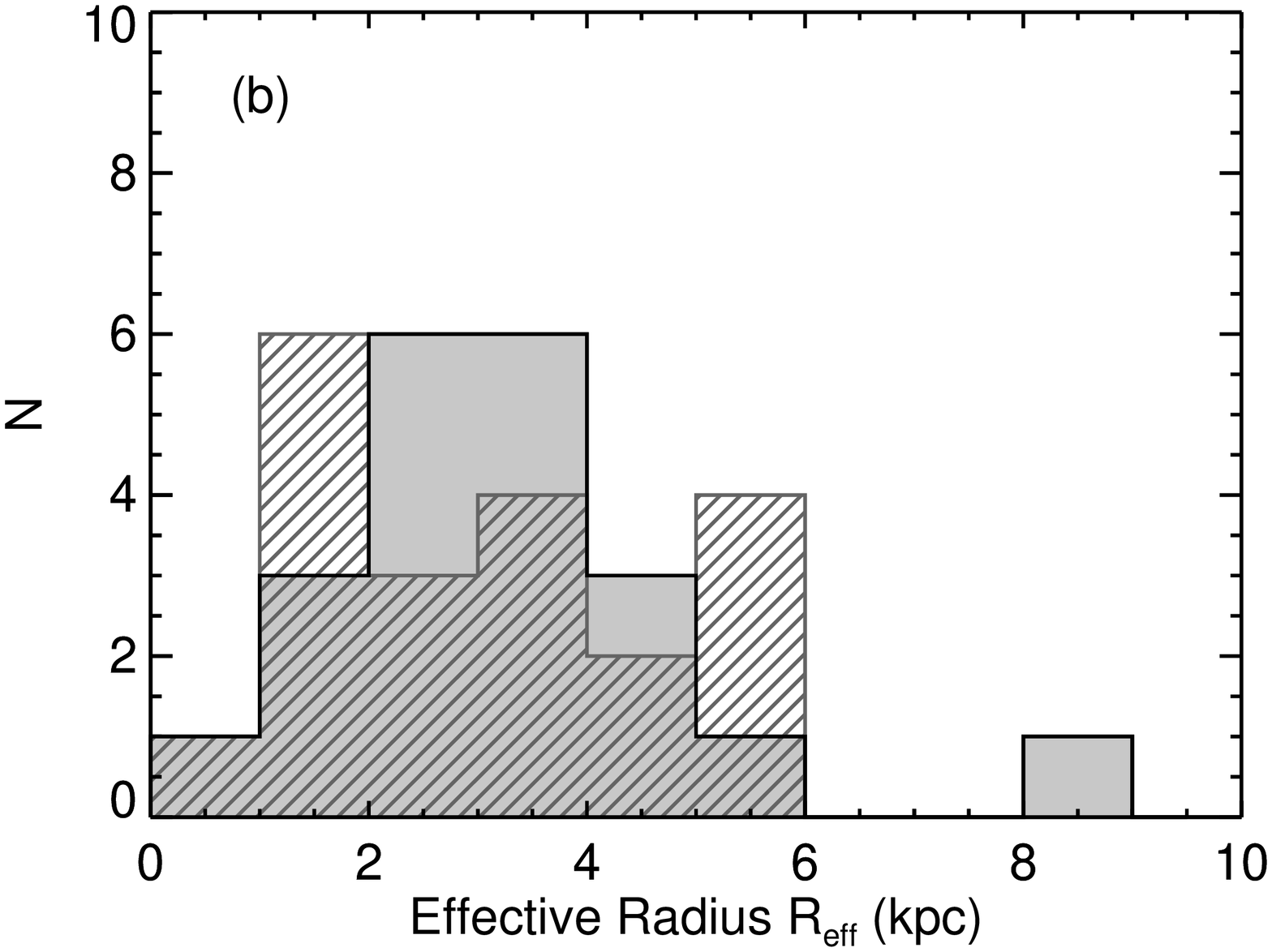}
\caption{Morphologies and radii of AGN host and matched comparison galaxies. \textit{Left:} the distribution of the of \sersic\ indices $n$. \textit{Right:} the physical effective radii $R_{\rm eff}$. The solid histogram represents the X-ray selected AGN host galaxies at $z\sim2$, while the striped histograms represent a matched comparison sample of normal galaxies. According to a Kolmogorov-Smirnov (KS) test,  the distributions of \sersic\ indices and effective radii of the AGN host galaxies and inactive comparison galaxies are consistent with being drawn from the same parent distribution. \label{fig:results}}

\end{center}
\end{figure*}


\section{Analysis}
\label{sec:analysis}

\subsection{Image Analysis with \galfit}
We used the two-dimensional fitting algorithm \galfit\ \citep{2002AJ....124..266P} to quantitatively analyze the light distribution of all X-ray selected AGN host galaxies. We ran \galfit\ on the $H$-band image cutout and associated noise map for each object. We manually masked all other sources in the cutout by setting the pixel values of the noise map to $10^{9}$, thus giving them no weight in the fit. We built an empirical point spread function (PSF) from a set of bright stars in the ERS field (see Figure \ref{fig:gallery}, bottom-right). Each star included in the empirical PSF was visually inspected for any possible contaminants from nearby stars. We checked that no star was under- or over-exposed and that the final composite PSF fits well the individual stars regardless of their spectral energy distribution. 

For each object, we fit three different models of the brightness distribution: (1) \textbf{PSF-only fit},to determine whether the source is resolved; (2) \textbf{\sersic-only fit} \citep{1968adga.book.....S}, to measure the \sersic\footnote{A value of \sersic\ index $n=1$ corresponds to a disk and $n=4$ to an elliptical.} $n$ and the effective radius $r_{\rm eff}$; and (3) \textbf{\sersic+PSF fit}, to represent an unobscured AGN point source.

The analysis of the two-dimensional light distribution of AGN host galaxies at high redshift poses two major challenges: the host galaxy may not be resolved and the host galaxy can be affected by the presence of a nuclear point source. First we determine which AGN host galaxies are resolved: following \cite{2009ApJ...705..639B} we compared the reduced $\chi^2$ of the \sersic\ and PSF fits to determine whether the \sersic\ fit is significantly better than the PSF fit, \textit{i.e.}, there is some extra light that can be better fit if we allow for an extended source. That is, we determine

\begin{equation}
F = \frac{(\chi^2_{\rm{PSF}} -  \chi^2_{\rm{Sersic}})}{\chi^2_{\rm{Sersic}}} > F_{\rm{crit}} ,
\end{equation}

\noindent where $F_{\rm{crit}}$ is the critical value at which the \sersic\ fit is significantly better. As \cite{2009ApJ...705..639B} argue, the value of $F_{\rm{crit}}$ cannot be determined from the F-distribution due to the behavior of \galfit\ near minima and so $F_{\rm{crit}}$ should be determine empirically. 

We therefore performed both \sersic\ and PSF fits on a set of 18 faint ($R > 23$ AB) stars in the ERS field and calculated $F$ for each. The values of $F$ for these stars ranges from $F \sim- 1$, where the \sersic\ fit converges to an unphysically small radius and high $n$, to cases where $F$ approaches $\sim0$. The highest value we find is $F=-0.001$; to be safe, we adopt $F_{\rm crit} = 0.01$ as \cite{2009ApJ...705..639B} suggest. 

This test removes 2 out of 23 AGN host galaxies. We also remove one source (ID\# 56769) that is too faint in the F160W image (see \citealt{2004ApJ...600L.123K} who first described it; it is a real source and not a spurious X-ray detection). One of the resolved sources, ID\# 58330, required two components to adequately fit, which we from now on refer to as components A and B.  We are left with 20 resolved AGN host galaxies. We also performed the \sersic-only fit to a set of comparison galaxies matched in both $R$-band luminosity and redshift. We show an example fit in Figure \ref{fig:examples}.

\begin{deluxetable*}{lcccccccccl}
\tablecolumns{11}
\tablewidth{0pc}
\tabletypesize{\normalsize}
\tablecaption{\galfit\ Results for X-ray selected AGN Host Galaxies at $z\sim2$}
\tablehead{
 \colhead{} & 
 \colhead{} & 
 \multicolumn{3}{c}{\small{\sersic\ fit only}} & 
 \colhead{} & 
 \multicolumn{3}{c}{\small{\sersic\ + PSF fit}} & 
 \colhead{} \\
 \colhead{ID$^1$} &
 \vline &
 \colhead{\sersic\ index $n$} &
 \colhead{Angular $r_{\rm{eff}}$} &
 \colhead{Physical $R_{\rm{eff}}$} &
 \vline &
 \colhead{\sersic\ index $n$} &
 \colhead{Angular $r_{\rm{eff}}$} &
 \colhead{Physical $R_{\rm{eff}}$} &
 \colhead{$\Delta m_{\rm AGN}$}  \\
 \colhead{} & 
 \colhead{} & 
 \colhead{} & 
 \colhead{\arcsec} & 
 \colhead{$kpc$} & 
 \colhead{} & 
 \colhead{} & 
 \colhead{\arcsec} & 
 \colhead{kpc} & 
 \colhead{mag} & 
 \colhead{} 
}
\startdata
\cutinhead{Galaxy-dominated, $\Delta m_{\rm AGN} = m_{\rm gal} - m_{\rm PSF} < -1$}\\
49190& & $         0.52 \pm         0.04$ & $        0.39 \pm         0.05$ & $        3.15 \pm         0.41 $ & & $        0.06 \pm         0.09$ & $        0.46 \pm         0.06$ & $        3.74 \pm         0.49$ &        -1.93 &   \\
50057& & $        10.44 \pm         2.33$ & $        0.10 \pm         0.01$ & $        0.78 \pm         0.10 $ & & $       14.03 \pm         5.09$ & $        0.07 \pm         0.01$ & $        0.55 \pm         0.07$ &        -3.54 &   \\
52141& & $         1.65 \pm         0.03$ & $        0.30 \pm         0.04$ & $        2.53 \pm         0.33 $ & & $        1.91 \pm         0.04$ & $        0.31 \pm         0.04$ & $        2.65 \pm         0.34$ &        -2.62 &   \\
52399& & $         1.29 \pm         0.11$ & $        0.23 \pm         0.03$ & $        1.99 \pm         0.26 $ & & $        0.31 \pm         0.10$ & $        0.34 \pm         0.04$ & $        2.90 \pm         0.38$ &        -1.12 &   \\
53849& & $         3.52 \pm         0.08$ & $        0.16 \pm         0.02$ & $        1.32 \pm         0.17 $ & & $        1.67 \pm         0.08$ & $        0.23 \pm         0.03$ & $        1.96 \pm         0.25$ &        -1.24 &   \\
55062& & $         0.89 \pm         0.02$ & $        0.38 \pm         0.05$ & $        3.08 \pm         0.40 $ & & $        0.35 \pm         0.02$ & $        0.44 \pm         0.06$ & $        3.61 \pm         0.47$ &        -2.04 &   \\
55620& & $         2.37 \pm         0.09$ & $        0.25 \pm         0.03$ & $        2.03 \pm         0.26 $ & & $        2.48 \pm         0.09$ & $        0.27 \pm         0.04$ & $        2.24 \pm         0.29$ &        -2.91 &   \\
56112& & $         1.99 \pm         0.14$ & $        0.61 \pm         0.08$ & $        4.79 \pm         0.62 $ & & $        0.05 \pm         0.04$ & $        0.77 \pm         0.10$ & $        6.11 \pm         0.79$ &        -1.47 &   \\
57805& & $         8.33 \pm         0.27$ & $        0.40 \pm         0.05$ & $        3.39 \pm         0.44 $ & & $        7.75 \pm         0.23$ & $        0.35 \pm         0.05$ & $        2.98 \pm         0.39$ &        -4.90 &   \\
58039& & $         6.35 \pm         0.21$ & $        0.62 \pm         0.08$ & $        5.22 \pm         0.68 $ & & $        6.25 \pm         0.21$ & $        0.58 \pm         0.07$ & $        4.87 \pm         0.63$ &        -5.18 &   \\
58330A& & $         0.79 \pm         0.05$ & $        0.31 \pm         0.04$ & $        2.61 \pm         0.34 $ & & $        0.74 \pm         0.04$ & $        0.45 \pm         0.06$ & $        3.76 \pm         0.49$ &        -1.51 &    \\
58330B& & $         2.28 \pm         0.17$ & $        0.48 \pm         0.06$ & $        4.04 \pm         0.53 $ & & $        2.50 \pm         0.18$ & $        0.44 \pm         0.06$ & $        3.65 \pm         0.47$ &        -4.42 &  \\
59060& & $         0.55 \pm         0.24$ & $        0.29 \pm         0.04$ & $        2.37 \pm         0.31 $ & & $        0.70 \pm         0.07$ & $        0.26 \pm         0.03$ & $        2.12 \pm         0.28$ &        -3.46 &   \\
\cutinhead{AGN-dominated, $\Delta m_{\rm AGN} = m_{\rm gal} - m_{\rm PSF} > -1$  (\sersic-only fit may be unphysical))}\\
50333& & $         3.66 \pm         0.59$ & $        0.04 \pm         0.01$ & $        0.28 \pm         0.04 $ & & $        0.84 \pm         0.49$ & $        0.15 \pm         0.02$ & $        1.22 \pm         0.16$ &         0.95 &   \\
56954& & $         9.74 \pm         1.96$ & $        1.60 \pm         0.21$ & $       13.34 \pm         1.73 $ & & $        0.17 \pm         0.03$ & $        0.58 \pm         0.08$ & $        4.87 \pm         0.63$ &        -0.89 &   \\
57420& & $        20.00 \pm         8.11^{2}$ & $        4.71 \pm         0.61$ & $       39.87 \pm         5.18 $ & & $        7.52 \pm         0.47$ & $        0.36 \pm         0.05$ & $        3.06 \pm         0.40$ &        -0.92 &   \\
57859& & $         1.62 \pm         0.14$ & $        0.34 \pm         0.04$ & $        2.63 \pm         0.34 $ & & $        1.08 \pm         0.17$ & $        0.39 \pm         0.05$ & $        3.04 \pm         0.40$ &         0.73 &   \\
58224& & $         4.22 \pm         0.40$ & $        0.11 \pm         0.01$ & $        0.91 \pm         0.12 $ & & $        0.97 \pm         0.16$ & $        0.27 \pm         0.04$ & $        2.17 \pm         0.28$ &        -0.19 &   \\
58509& & $         1.57 \pm         0.18$ & $        0.14 \pm         0.02$ & $        1.22 \pm         0.16 $ & & $        0.06 \pm         0.39$ & $        0.30 \pm         0.04$ & $        2.57 \pm         0.33$ &        -0.23 &   \\
58657& & $         6.38 \pm         3.16$ & $        0.03 \pm         0.00$ & $        0.24 \pm         0.03 $ & & $        0.05 \pm         0.36$ & $        0.45 \pm         0.06$ & $        3.65 \pm         0.47$ &         1.44 &   \\
63732& & $        20.00 \pm         4.33^{2}$ & $        7.68 \pm         1.00$ & $       64.95 \pm         8.44 $ & & $        0.97 \pm         0.05$ & $        1.00 \pm         0.13$ & $        8.43 \pm         1.09$ &        -0.55 &   \\
\enddata

\label{tab:galfit}

\tablenotetext{1}{Catalog ID, see Cardamone et al. (2010).}
\tablenotetext{2}{The maximum value of the \sersic\ index allowed is $n=20$, so fits with $n=20$ did not reach an acceptable minimum.}

\end{deluxetable*}

\section{Results}
\label{sec:results}
In Table \ref{tab:galfit}, we report the results for all 20 AGN host galaxies\footnote{21 objects, if 58330A and B are counted individually.} that are well-resolved. For the final analysis, we needed to decide which fit to use. We therefore divide our sample into galaxy-dominated objects (lacking a strong nuclear point source) and AGN-dominated objects using the magnitude difference of the host galaxy to the AGN from the \sersic+PSF fit. Specifically, we use  the magnitude difference $\Delta m_{\rm AGN} = m_{\rm gal} - m_{\rm PSF} = -1$ as the partition (varying this does not significantly change the results). 
In what follows we quote the best \sersic-only fit when the galaxy is at least 1 magnitude brighter than the point source, and the best \sersic+PSF fit otherwise; we account for the two components of source ID\# 58330 separately. As discussed below, most fits favor low \sersic\ indices, so for all fits with $n<4$, we checked explicitly which fit better, a pure disk ($n=1$, fixed) or a pure bulge ($n=4$, fixed). In every case, the $n=1$ fit has a lower $\chi^2$ value. The systematics of decomposing nuclear point sources and extended host galaxies is discussed extensively by \cite{2008ApJ...683..644S} who show that as long as the point source is no more than four times brighter than the host galaxy, the fitted host galaxy parameters are reliable. In our sample, only one source comes close to this limit.

\subsection{AGN Host Galaxies at $z\sim2$}
We show the distribution of \sersic\ indices in Figure \ref{fig:results}a. The distribution is bimodal, with four AGN host galaxies having high \sersic\ indices of $n>4$ while the remaining population has low \sersic\ indices. The mean and median \sersic\ indices are $n=$2.54 and $n=$1.08, respectively. Most $z\sim2$ AGN host galaxies ($16/20 \sim 80\%$) thus have disk-like light profiles while only a minority appear to be dominated by bulges (Note that simulations show that at high redshift, disks and bulges can be confused, but for $n<3$, at least half of the light is likely from the disk  \citealt{2008ApJ...683..644S}). The distribution of disk vs. spheroid morphology is qualitatively similar to that of the local AGN host galaxy population of $\sim10\%$ early-type, and $\sim90\%$ late-types and indeterminate-type (\textit{i.e.,} Sa/S0) galaxies \citep{2010ApJ...711..284S}.

We show the distribution of effective radii in Figure \ref{fig:results}b. The distribution is unimodal and the mean and median effective radii are 3.16 kpc and 3.04 kpc, respectively. We compare these effective radii to the AGN host galaxy sample at $z\sim0.05$ presented by \cite{2010ApJ...711..284S}: the mean and median effective $g$-band\footnote{The $g$-band is the closest band compared to the rest-frame wavelength of the $F160W$ $H$-band filter.} radii of this sample are 3.17 and 2.84 kpc, respectively, very similar to the high redshift AGN. 

Figure \ref{fig:results} show the histograms for the \sersic\ indices and effective radii of the matched comparison sample in Figure \ref{fig:results} and perform a Kolmogorov-Smirnov (KS) test shows no evidence for the two populations being different; they are consistent with being drawn from the same parent distributions. 

\subsection{Black Hole Mass, Eddington Ratio and Dark Matter Halo Mass}
In order to give some context to the AGN in our sample, we consider their black hole and dark matter halo masses. The majority of our objects are disk-dominated so any black hole -- stellar mass relation can only be taken as a general guide. From the stellar mass estimates described in \cite{2010ApJ...721L..38C} using the FAST algorithm \citep{2009ApJ...700..221K}, we find that the AGN host galaxies in our sample have mean and median stellar masses of $10^{10}$\Msun and $2.5 \times 10^{10}$\Msun, respectively.  Using the black hole -- bulge relation of \cite{2004ApJ...604L..89H}, this should yield typical black hole masses of $5\times10^{7}$\Msun. Given that this relation does not account for the mass of the disk which is included in our stellar mass estimate, this is an upper limit. 

We calculate the Eddington ratios of our sources using the X-ray luminosity assuming an X-ray to bolometric correction of 20 and black hole masses, yielding mean and median Eddington ratios of 0.37 and 0.09, respectively (Table \ref{tab:agn}). Of course, the black hole mass is likely smaller and thus the Eddington ratio is higher so this probably a period of significant mass growth. This is commensurate with the Eddington ratios of broad-line AGN of comparable X-ray luminosity at this redshift of $\sim0.02-1$ with the bulk at $\sim0.3$ \citep{2010ApJ...708..137M, 2009ApJ...700...49T}. 

Finally, studies of the clustering of similar of X-ray AGN \citep[e.g.,][]{2008ApJ...673L..13F, 2009A&A...494...33G} indicate that they reside in high-mass dark matter haloes and that their $z=0$ descendants are massive galaxies, most likely massive early-type galaxies with massive black holes. This means that these AGN have grown into some of the most massive black holes locally.

\section{Discussion \& Conclusions}
\label{sec:discussion}

We have obtained the first clear view of the rest-frame optical morphologies of  AGN host galaxies with Seyfert-like luminosities ($10^{42} < L_{\rm X} < 10^{44}$\ergs) at $z\sim2$ using the new \textit{Hubble Space Telescope} WFC3/IR in the ERS portion of the GOODS-S field. Fits to the host galaxy surface brightness profiles reveals that:

\begin{enumerate}
\item The majority of these AGN host galaxies have low \sersic\ indices, implying the bulk of the host galaxy light comes from a disk.
\item The host galaxy structural parameters (\sersic\ index and effective radius) do not appear to be significantly different from a comparison sample of inactive galaxies matched in redshift and luminosity.
\item The distribution of \sersic\ indices implies that high redshift AGN host galaxies have very similar morphologies to local AGN host galaxies, \textit{i.e.},  few early types but many late types. The typical effective radii are also similar to those of local AGN host galaxies.
\end{enumerate}

These AGN host galaxies are a significant fraction of the total AGN population by number density and in terms of light emitted by accretion \citep[][]{2003ApJ...598..886U, 2005A&A...441..417H}. Using the X-ray luminosity function of \cite{2003ApJ...598..886U} and evolution, obscuration distribution and bolometric correction as described by \cite{2009ApJ...696..110T}, we estimate the black hole growth in this population in the $z=1.5-3$ range spanned by our sample represents 10--17\% of the total black hole growth over cosmic history. Excluding the most massive black holes, which get most of their mass in quasar-luminosity events triggered by mergers \citep{2010Sci...328..600T}, 23--40\% of black hole growth occurs in a secular mode driven by internal processes in the host galaxy. Since disks also dominate the AGN host galaxy population at $z\sim0$ \citep{2010ApJ...711..284S}, where quasar-mode growth is unimportant, an even larger fraction of \textit{all} black hole growth over cosmic history appears to take place in disk galaxies. 

The results presented here show that moderate luminosity AGN host galaxies at $z\sim2$ and $z\sim0$ are remarkably similar. The high fraction of AGN host galaxies with disk-like light profiles is difficult to reconcile with the expectation of black holes growing jointly with stellar bulges during special phases of their evolution, such as major mergers envisioned in many simulations \citep[e.g.,][]{2005MNRAS.361..776S, 2005ApJ...630..705H, 2008ApJS..175..356H}. The disk morphologies of the host galaxies point instead to secular processes \citep[e.g.,][]{2004ARA&A..42..603K} as most common growth mode. The fact that AGN host galaxies are indistiguishable from the  $z\sim2$ comparison sample in terms of their \sersic\ indices and effective radii further supports the role of secular growth. This is very different from the high-luminosity (quasar) population at the same redshift, which does seem to be driven by major mergers \citep{2010Sci...328..600T}. Thus in the high redshift universe, there appear to be two distinctly different modes of black hole growth for high- and low-luminosity AGN.

The fact that the majority of black hole growth in this population --- and by extension a significant fraction of cosmic black hole growth --- occurs in a galaxy substantial disk means that it is not associated with major mergers. This raises interesting questions regarding the origin and relevance of the relationship between galaxy and black hole mass \citep{2000ApJ...539L..13G, 2000ApJ...539L...9F, 2007ApJ...671.1098P, 2010arXiv1006.0482J}. This secular black hole growth must still be self-regulated in some way that preserves the correlation between black hole mass and bulge mass.


\acknowledgements 

Support for the work of KS and ET was provided by NASA through Einstein/Chandra Postdoctoral Fellowship grant numbers PF9-00069 and PF8- 90055, respectively, issued by the Chandra X-ray Observatory Center, which is operated by the Smithsonian Astrophysical Observatory for and on behalf of NASA under contract NAS8-03060. CMU and CC acknowledge support from NSF grants AST-0407295, AST-0449678, AST-0807570 and Yale University. SKY acknowledges the support by the National Research Foundation of Korea  through the Doyak grant (No. 20090078756) and the SRC grant to the Center for Galaxy Evolution Research. KS is grateful for the hospitality of the Astronomy Department at Yonsei University, Korea and thanks Yun-Kyeong Sheen, Sang-Il Han and Andrew Fruchter for technical help. This research has made use of NASA's Astrophysics Data System Bibliographic Services. \\
{\it Facility:} \facility{HST (WFC3)}, \facility{Chandra (ACIS)}

\bibliographystyle{apj}


\end{document}